# Data Flows and Colonial Regimes in Africa: A Critical Analysis of the Colonial Futurities Embedded in AI Recommendation Algorithms

Angella Ndaka,
University of Witwatersrand, Johannesburg, South Africa
Fátima Ávila-Acosta,
Berlin Graduate School of Social Sciences at Humboldt University,
Berlin, Germany
Harnred Mbula,
Centre for Epistemic Justice, Nairobi, Kenya
Christine Amera,
Centre for Epistemic Justice, Nairobi Kenya
Sandra Tiyani Chauke
University of Pretoria, South Africa
Eucabeth Majiwa
Jomo Kenyatta University of Agriculture and Technology, Nairobi, Kenya

**Abstract**
In the last few years, Africa has experienced growth in a thriving ecosystem of Artificial Intelligence (AI) technologies and systems, developed and promoted by both local and global technology players. While the sociotechnical imaginaries about these systems promote AI as critical to achieving Africa's sustainable development agenda, some of them have subtly permeated society, recreating new values, cultures, practices, and histories that threaten to marginalize minority groups in the region. Africa predominantly frames AI as an imaginary solution to address complex social challenges; however, the narrative subtly ignores deeper power-related concerns, including data governance, embedded algorithmic colonialism, and the exploitation that propagates new digital colonial sites. However, the development of current AI ethics in Africa is in its infancy and predominantly framed through lenses of Western perspective, with the social and ethical impacts of the AI innovations and application on African epistemologies and worldviews not prioritized. To ensure that people on the African continent leverage the benefits of AI, these social and ethical impacts of AI need to be critically and explicitly considered and addressed. This chapter will therefore seek to frame the elemental and invisible problems of AI and big data in the African context by examining digital sites and infrastructure through the lens of power and interests. It will present reflections on how these sites are using AI recommendation algorithms to recreate new digital societies in the region, how they have the potential to propagate algorithmic colonialism and negative gender norms, and what this means for the regional sustainable development agenda. The chapter proposes adopting business models that embrace response-ability and consider the existence of alternative socio-material worlds of AI. These reflections will mainly come from ongoing discussions with Kenyan social media users in this author's user space talks, which take place every month.

**Section 1: Introduction**

The growing global interest, combined with rising investments in AI skilling and infrastructure development, is a key driver of the expanding landscape of AI technologies and systems across Africa. This interest has not only led to increased engagement among many Africans in conversational AI, but also to an emerging plethora of grassroots initiatives that are catalysing capacity building, local solutions, and an AI knowledge base across the continent. For instance, the Stanford AI Index (2024) showed that 24% of the Kenyan population used OpenAI ChatGPT daily, and 2025 statistics placed Kenya at the top of the global map, with over 42% using ChatGPT daily. In the continent's markets and professional settings, Google AI- related searches have increased by over 270% in a year, highlighting increasing curiosity among the public and professionals (West 2024). At the same time, there is an emergence of local initiatives, whose agenda is to drive African developed and owned AI solutions and knowledge, e.g., Deep learning Indaba, Data science Africa, Ushahidi, and Tayarisha centre, Wits MINDS among others who are not only catalysing research-led AI development and capacity building, but are also using AI for civic advocacy across the continent, strengthening the AI knowledge co-production base (Brookings 2024), as well as exploring the AI and digital tools to enable economic and political agency, response-able digitization, and digital resilience in the continent (Ndaka, Oando & Majiwa 2024). While local innovations' scaling is minimal due to critical under-resourcing, there is anticipation that AI may contribute over $1.2 billion to the continent's GDP by 2030, across major sectors (Agrifocus Africa 2024). This growth in GDP will come from increased AI adoption across sectors like healthcare, education, public services and agriculture, as well as a growing diversity of startups and AI solutions in those sectors.

Despite local actors showing resilience in their quest to demonstrate Africa's potential to build and meet the demand for context-aware AI solutions (Zimba et al., 2025), which resists an import and consumer-driven approach to AI (Ndaka 2024), there are significant gaps in addressing power dynamics. Global technology multinationals like Google, Microsoft, Meta, OpenAI, and IBM have firmly entrenched themselves in the African AI ecosystem through investments in AI research relevant to them, skilling, and talent incubation (Brookings, 2024), while dominating AI debates. While these activities appear as positive investments made by big tech, the invisible activities running through the business models, e.g., online surveillance, recommendation algorithms, data scrapping, and labour extraction through proxies (BPOs) should be of concern – as they are entrenching colonial imprint through data extraction, transposed digital infrastructure and commodification of everyday social interaction through social digital spaces (Ndaka et al 2024). Worse, the government's involvement in the hype creates a significant conflict of interest between its role as custodian of citizens and private interests (Ndaka 2017). This emphasizes what many scholars have argued in the past about technology and its investment – as a political tool that has enacted political goals for centuries (Latour 2011; Hecht 2009). These tools foreground dominant power structures, capitalist norms, and global corporate dominance (Fauset 2008). By design, they have prioritized foreign interests, compromising Africans' ability to assert their agency in socio-material relationships that shape how AI and emerging technologies unfold in the African continent (Ndaka, Oando & Majiwa 2025). Global North and Asian agencies are not only controlling vast amounts of data (Healy and Fourcade 2025) but are also patronizing the knowledge production and dissemination systems in the continent, albeit through African proxies (Ndaka et al 2024; Ndaka & Majiwa 2024; Harding 2001).

Major policy tools such as the SDGs and the Africa Agenda 2063 have increasingly framed AI as a development enabler, thereby centering discussions around AI development. The Continental AI strategy and multiple other national AI strategies have positioned AI as a force for Africa's economic growth and a catalyst for equity across sectors (African Union 2024). But is this feasible? Will AI save Africa from its wicked problems?. Ndaka (2024), in her work that focuses on how Africa is assembling policies in anticipation of AI-driven techno-futures, critically highlights how current policies foreground the interests of those in power, and, by design leave out the critical voices of institutions and people in the formulation and potentially implementation. This chapter will therefore seek to frame the elemental and invisible problems of AI and big data in the African context by examining commonly used social media and search sites; Facebook, Google, TikTok, Instagram, LinkedIn, and Twitter, through the lens of power and interests. It will present reflections on how these sites are using AI recommendation algorithms to recreate new digital societies in the region, how they have the potential to propagate algorithmic colonialism and negative gender norms, and what this means for the regional sustainable development agenda. This chapter proposes adopting business models that embrace response-ability and consider the existence of alternative socio-material worlds that apply AI. The reflections shared mainly come from the ongoing discussion with Kenyan social media users in this author's user space talks, which take place every month.

## The Coloniality in Digital Technologies

The rapid development and accelerated deployment of Artificial Intelligence (AI), algorithmic systems, and digital infrastructures worldwide are profoundly reshaping social, cultural, and political dynamics as we know them. These digital systems are neither abstract nor neutral (Hasselbalch 2021); instead, they are laced with political and corporate interests, while maximizing on invisible human contributions – labour and data - powering their emergence (Hasselbalch 2024). Developed mainly upon Western-centered values, these top-down technologies embody tangible structures that reflect the political and ideological interests of their developers (Salami, 2024; Ricaurte, 2019). Consequently, their widespread adoption carries profound and complex implications that must be critically reflected upon.

These automated computational systems are forms of power that operate under colonial logics. The use of these technologies has extended extractivist and exploitative practices from physical to digital territories (Gago, 2020; Mohamed et al., 2020). In these digital territories, algorithms have become the Western interlocutors of emerging social configurations and imaginaries. Subjected to "conquest" by Western values and capitalist logics, digital territories have become critical arenas for reinforcing colonial relations (Ricaurte, 2019).

With colonial practices perpetuating in the digital realm, expansionist regimes of power continue to operate and intensify through new abstract mechanisms that, while intangible, have profound effects on the lived realities of communities. Power relations rooted in colonialism manifest in multiple, interconnected, and layered dynamics within these systems (Ricaurte, 2019). Displayed through both material and digital forms of extraction and exploitation, these ongoing dynamics sustain broader dominant and oppressive structures, reinforcing conditions of dependency, concentrated wealth accumulation, and power asymmetries between the Global North and the Global South (Couldry & Mejias, 2019).

### *From the Physical to the Digital: The Expansion of Colonial Power Dynamics*

Resource extractivism and labor exploitation for digital industry operations primarily begin in physical territories. The construction of digital infrastructure, the operation of computational systems, and maintenance of data centers and servers—as well as other essential industry activities —demand vast natural resources. Resources often come from territories in the Global South, where intensive consumption causes environmental degradation and limits local communities' access to them . Although most data centers are in the Global North, these facilities are gradually expanding into the Global South in search of greater resource availability and lower operational costs.

Tech industries engage in exploitative labor practices by hiring workers from the Global South, where pervasive inequality, economic vulnerability, and weak labor laws enable them to minimize manufacturing costs by paying extremely low, unfair wages (Salami, 2024; Mohamed et al., 2020). The so-called "ghost workers", who perform critical yet invisibilized tasks such as data labelling and content moderation, frequently endure poor working conditions with inadequate protection despite being exposed to highly traumatic content (Gray and Suri, 2019).

The continuity of these colonial logics extends primarily, though not exclusively, into digital territories through data extractivist practices and algorithmic colonialism that often occur without transparency, user consent, or meaningful control (Couldry & Mejias, 2019). Data has become a valuable resource extracted from local users and transferred to geographically distant servers for processing, refinement, and synthesis (Coleman, 2019). The treated data is then used to develop new products, improve ad targeting, perform predictive analytics, and support other activities aimed at profit maximization (Coleman, 2019; Ndaka 2024). The derived economic benefits are not equitably shared with the communities where the data originates. Instead, wealth concentrates among a small number of tech elites, expanding their financial power and control, and deepening global inequalities (Coleman, 2019; Couldry & Mejias, 2019; Salami, 2024). This form of dispossession systematically excludes communities from control, access, and benefits derived from their data, which is owned, commodified, and exploited for monetary profit, which mirrors colonial logics as discussed by scholars (e.g., Ndaka et al., 2024; ….)

With the emergence of digital territories, algorithms have become the mediators of subjectivities and social relations in the digital realm. But algorithms are not neutral; they reflect—and consequently promote—dominant cultural norms, value systems, and practices. Algorithms are about power (Mohamed et al., 2020). Their design is not democratized and is rooted in a clear geopolitical standpoint; they lack context awareness and therefore do not represent everyone.

By visibilizing and amplifying dominant narratives and viewpoints, they neglect and marginalize non-Western epistemologies, cosmovisions, and alternative ways of living, actively harming underrepresented and minority groups (Ricaurte, 2019). This distorted homogenization of digital territories promotes a false universality that erases plurality and replicates colonial social hierarchies (Nothias, 2025)—ultimately, functioning as a mechanism of dispossession that uproots cultural identities and displaces traditional epistemological systems and indigenous knowledges (Ricaurte, 2019). Algorithmic narratives impact not only cultural identities but also shape political and economic structures.

Together, these material and digital forms of extraction and exploitation sustain and perpetuate broader colonial dynamics by concentrating wealth and power unevenly (Ricaurte, 2019). The benefits are neither shared fairly, redistributed, nor compensated to the countries being exploited (Coleman, 2019). Instead, resources continue to accumulate mainly in the Global North (Coleman, 2019). This concentration of profit, along with dominance and control over digital technologies and infrastructure, create power imbalances that lead to dependence on foreign tech companies (Salami, 2024; Nothias, 2025).

In addition, the lack of ownership rights and agency over locally produced datasets reduces national autonomy and limits sovereignty (Coleman, 2019). By dispossessing users of agency, consent, and control over their data, these technologies not only strip them of their dignity as data producers but also deprive them control of their actions or choices, with socio-material consequences (Ndaka, Oando & Majiwa 2025). This fundamental loss of autonomy and dignity may seem radical, but it is real and acts as a key mechanism in replicating and intensifying racial disparities embedded in digital technologies. Consequently, their use and inevitable expansion profoundly impact communities, disproportionately affecting those already marginalized.

Furthermore, these mechanisms are disseminated behind a solutionist discourse and framed under the classical colonial narrative of progress and development (Mohamed et al., 2020). Computational systems and AI technologies are promoted under the guise of a universal one-size-fits-all solution indispensable to progress, masking their underlying and harmful power dynamics (Coleman, 2019). This discourse normalizes digital colonial practices, which have been accepted as the *status quo* (Couldry & Mejias, 2019), thereby perpetuating coloniality in structural and unchallenged ways and expanding it into digital territories.

*The Socio-Technical Perpetuation and Reinforcement of Colonial Logics*

There is no doubt about the promising potential of AI and digital technologies. However, a critical reflection on the implications of their adoption is urgent. Coloniality is alive in the socio-technical reality we live in and is being reinforced through extractivist and exploitative practices that have expanded from the physical to the digital (Coleman, 2019). The colonial geographical hierarchies remain very pronounced when it comes to the sources of digital labor and raw materials for the digital industry. These infrastructures are being built upon historic colonial legacies and are being developed at the expense of communities; ultimately deepening global inequalities by reproducing systems of oppression and power disparities (Mohamed et al., 2020).

Computational technologies embody clear capitalist ideologies and expansionist power systems. Framed as a fiction of progress, digital systems and algorithms serve as tools for economic, social, and political control with a bias toward the interests and norms of dominant powers (Ricaurte, 2019). Furthermore, there is a dangerous normalization of the unfair, unequal, and undignified dispossession of ownership and sovereignty over data, paving the road for the unrestrained dominance of tech elites (Couldry & Mejias, 2019; Ricaurte, 2019).

**Moving Ethics from Paper to Practice**

In Africa, AI is often framed by AI optimists as a silver bullet: predictive agriculture to combat food insecurity, digital finance to close the unbanked gap, or algorithmic healthcare to expand access in underserved communities. These promises are wrapped in the language of opportunity and inclusion, suggesting that the continent can leapfrog into the Fourth Industrial Revolution. Yet the lived reality is far more complicated.

While *ethics in AI* is frequently mentioned in policy documents, corporate statements, and international frameworks, its actual practice remains elusive. Too often, African people are subjected to algorithms built elsewhere, with values and assumptions that do not reflect their lived experiences. Instead of inclusive progress, AI risks deepening inequalities, silencing local epistemologies, and repeating the extractive logics of colonialism - only this time through data and algorithms. The gap between paper promises and lived realities raises a pressing question: whose ethics, whose values, and whose interests guide the deployment of AI in Africa?

The central argument of this chapter is that ethics in AI for Africa cannot remain rhetorical. It must be embedded in everyday governance, grounded in African philosophies, and responsive to local contexts. Without this shift, AI risks reproducing the extractive logics of colonialism through data, algorithms, and digital infrastructures. This discussion builds on scholarly work such as *Trustworthy AI: African Perspectives* (Oguamanam et al., 2022), which emphasizes that ethical AI must be situated, contextual, and rooted in African lived realities rather than imported wholesale from Western paradigms.

In Africa, it is difficult to separate AI from broader histories of power. Just as colonialism justified itself through development narratives while extracting land and labor, today's data economy is framed as a pathway to digital inclusion while replicating asymmetries. Nick Couldry and Ulises Mejias describe this process as "data colonialism," where human behavior is appropriated as raw material (Couldry & Mejias, 2019). For African users, this often means that simply logging into global platforms like Facebook or TikTok contributes to systems that commodify preferences and emotions, mainly for the benefit of corporations headquartered in the Global North.

Abeba Birhane (2020) extends this critique in her work on algorithmic colonialism, noting how systems designed for Western contexts are deployed uncritically in Africa, erasing cultural nuance and reinforcing dependency. The most visible example is in labor-sustaining AI systems. Content moderators and data annotators in Kenya and Uganda provide the hidden labor behind Big Tech's algorithms, often for meager wages, while bearing the psychological cost of reviewing traumatic material (Muldoon et al., 2023). Their contribution is indispensable, yet their dignity and well-being remain peripheral to global debates on AI ethics. The continuity with earlier colonial patterns is clear: extraction, invisibilities, and asymmetrical benefit.

Beyond labor exploitation, biased AI systems deepen inequality across critical sectors. In healthcare, algorithms trained on datasets from the Global North often fail to capture African genetic, environmental, and epidemiological realities. Obermeyer et al. (2019) demonstrated how an algorithm in the United States underestimated illness severity in Black patients because it used healthcare costs as a proxy for need. Transposed into African contexts, such flawed logic would exclude the poor, who spend less on healthcare despite a greater need for essential services.

Language bias offers another example of exclusion. Africa is home to more than 2,000 languages, yet most natural language processing systems prioritize English, French, and other colonial languages. This not only marginalizes speakers of African languages but also entrenches epistemic hierarchies where what is not legible to machines is treated as invisible. The grassroots Masakhane project has demonstrated that African researchers can build open, community-led language models (Nkwazi Magazine, 2022). Still, without sustained investment, linguistic inequity risks persisting in the digital age.

Surveillance technologies compound these risks. Facial recognition systems deployed in African cities frequently misidentify darker-skinned individuals, particularly women (Buolamwini & Gebru, 2018). In contexts where legal safeguards are weak, such errors translate into wrongful arrests, heightened surveillance, and suppression of dissent. When bias intersects with authoritarian tendencies, the ethical stakes of AI become a matter of survival rather than abstraction.

Governments across Africa have taken steps to craft AI strategies and policy frameworks. The African Union has circulated drafts, while countries such as South Africa, Rwanda, Kenya, and Nigeria have developed national development strategies . However, most policies echo Western ethical guidelines - fairness, accountability and transparency, without adaptation to local contexts (CIPIT, 2023). The result is often ethics as rhetoric, with little implementation. The 2024 Oxford Insights AI Readiness Index ranked Africa lowest globally, citing weak infrastructure, limited skilled capacity, and fragile institutions (Oxford Insights, 2024). But beyond capacity constraints lies a deeper issue: ethics is inconvenient. Corporations avoid rigorous ethical commitments because they slow innovation and cut into profits. Governments, dependent on foreign investment, hesitate to regulate Big Tech too strongly. The failure to regulate produces what Ndaka (2024) terms "compromised agency," in which African policymakers, workers, and even scholars become constrained from voicing critiques for fear of losing opportunities or livelihoods. Ethics thus becomes aspirational language without enforcement mechanisms.

Ethics without enforcement is ineffectual. Africa needs anticipatory and protective regulatory frameworks grounded in a duty of care. This means governments, corporations, and developers are legally and morally obliged to safeguard the well-being of those affected by AI. Practical steps include, ethical auditing and certification of AI systems before deployment, worker protections for annotators and tech laborers, including fair wages and mental health support, data sovereignty laws ensuring Africans control how their data is collected, stored, and used, community participation in AI governance, especially for marginalized groups and transparency obligations for corporations deploying AI in sensitive sectors like healthcare or policing.

Ethics in AI goes beyond embedding ethics in law, education, and institutional practice. It entails participatory design methods, where communities co-create AI tools with developers to ensure that AI systems reflect real needs rather than imposed assumptions. It also involves dialogue platforms with African social media users, as piloted in Kenya, which demonstrated that community concerns ranging from online safety to exploitative working conditions meaningfully inform governance structures (*Trustworthy AI: African Perspectives*, 2025).

For Africa, ethical AI is not an optional add-on. It is a matter of survival. Without it, the continent risks sliding into digital colonialism, with data extraction and algorithmic decision-making reinforcing inequality and undermining sovereignty. With it, AI can become a tool of liberation, enabling communities to thrive in dignity and justice.

As the contributors to *Trustworthy AI: African Perspectives* argue, the time for aspirational language is over. Ethics must move from paper to practice, from the margins to the center, from rhetoric to lived realities. That requires courage: from governments to regulate, from workers to speak, from communities to demand, and from scholars to imagine alternatives. The future of AI in Africa depends on it.

**Data Governance and Algorithmic Colonialism**

In the 21st century, data has become one of the most valuable resources in the global economy, often described as the 'new oil'. This comparison highlights its economic significance and points to the risks of data extraction, dependency, and unequal benefit distribution. Africa is bound to the legacies of resource extraction, underscoring the urgency. Africa's data is harvested, stored, and monetized by external actors. Data governance is the system of rules, policies, and practices that manage data. Data governance can and will enforce control, transparency, and local empowerment. However, the global landscape of AI and data governance shows deep inequalities; data and algorithms are not neutral. They are shaped by specific cultural, political, and economic contexts, often influenced by the global north, leading to algorithmic colonialism. Algorithmic colonialism allows technological systems to reinforce historical power imbalances, extracting value from African societies while leaving little room for local control (Birhane, 2020; Couldry & Mejias, 2019). Focusing on Africa, the decision-making algorithms that underpin an AI system do not lie within the continent but elsewhere. The people who built these systems, primarily from the global north, frame Africa to align with their ideologies and biases.

This chapter highlights the emergence of new digital colonial sites in Africa, the power asymmetries, embedded algorithmic colonialism, and exploitative practices, as well as the urgent need for localized frameworks for data sovereignty.

The emergence of new digital colonial sites in Africa is a fascinating yet concerning trend. The rise of digital platforms has significantly transformed African societies, opening the door for innovation, entrepreneurship, and connectivity. For instance, platforms like Google and Facebook collect vast amounts of user data from the Global South, which they monetize to train algorithms and develop AI products. While profits overwhelmingly benefit corporations in Silicon Valley, countries that provide data often see little to no return ( Couldry & Mejias, 2019).  Digital data is collected locally but is processed and stored in data servers outside Africa. Many African countries rely heavily on foreign-owned data centers and cloud services, e.g, Google Cloud and Microsoft Azure. Africa's internet connectivity depends on undersea cables financed and controlled by international corporations, leaving it reliant on external entities for essential infrastructure (Mwema et al., 2024). The reliance on external frameworks weakens sovereignty, urgency, and self-determination; hence, there is extensive mimicking of GDPs and strategy frameworks from the global north. This reliance creates structural dependency in which Africa consumes digital products and services but plays a limited role in shaping their design.

AI mirrors and reinforces most of the existing global inequalities. Most AI systems are trained Most AI systems are trained primarily on models from and for the Global North, often reflecting Western values, norms, languages, and assumptions that frequently marginalize the voices of subalterns. Data ownership is primarily concentrated among global tech firms, leaving the question of who owns the data - and who controls the data that controls us. Most AI research and patents are regulated and managed by big tech corporations and elite universities, such as Google and Meta, which not only lead in technology but also shape the narrative of what AI should do and to whom. African stakeholders often remain consumers, data providers, or test subjects through social media or participation in digital labor platforms rather than co-creators of AI systems, thereby reproducing knowledge hierarchies in African AI systems. African nations are struggling with limited resources and exclusion from major AI developments and research communities, leading to significant power imbalances. As Birhane (2020) points out, AI becomes another way to reinforce colonial hierarchies and marginalization of African epistemologies and ways of understanding.

The exploitative practices of algorithmic colonialism in Africa go beyond the power asymmetries, leading to real harm. Dominant AI systems reflect masculine world views and Western culture, causing algorithmic bias and discrimination. Facial recognition systems often misclassify darker skin tones at significantly higher rates than lighter ones (Boulamwini & Gebru, 2018). These technological errors are not accidental; they result from training models and processes that systematically ignore African populations and favor Westernized processes, leading to racial vulnerabilities, especially in surveillance and policing. Additionally, AI development often relies on outsourced labor, primarily from underpaid workers in Africa and postcolonial regions such as India. The scenario mirrors colonial systems of cheap labor exploitation (Gray & Suri, 2019). Algorithmic colonialism also includes environmental exploitation. Training large AI models requires substantial energy and resources, thereby imposing hidden environmental costs on foreign data centers and the AI economy. Additionally, digital platforms set content moderation policies that often silence African voices or misinterpret African narratives, reinforcing stereotypes instead of showcasing diverse perspectives. Embedded algorithmic colonialism is not just a technological issue but a deeply political and economic issue, reflecting repeated historical patterns of extraction and inequality.

One of the biggest challenges in Africa is the lack of a strong, localized system for data governance and sovereignty. While a few African nations have made progress, such as Nigeria's Data Protection Regulation (NDPR) and South Africa's Protection of Personal Information Act (POPIA), the overall regulatory framework remains fragmented and underdeveloped. The African Union has started to address these gaps with its data policy framework (2022). This framework emphasizes data sovereignty and regional collaboration. However, implementation varies across nations, and enforcement of these frameworks is limited.

Most African countries lack localized infrastructure, thus weakening data sovereignty. With most African data processed and managed overseas, under foreign jurisdictions and policies, this raises sovereignty and privacy issues and increases dependency risk, preventing African nations from comprehensively governing their own digital future.

Localized frameworks are urgently needed to ensure African data is not only protected but also used in ways that benefit local communities. Data sovereignty is not only about regulation but also about empowering African nations with the authority, skills, and resources to govern their digital systems.

**AI Recommendations Algorithms within the context of social media**

Artificial intelligence powers algorithms, which underpin the entire ecosystem of how social media platforms operate. These platforms can sort posts, videos, and ads into what are deemed personalised feeds, which shape the content people are exposed to and, thus, public discourse. Narayanan (2023) explains that these recommender systems can decide who sees which posts and are central to platforms such as Facebook, TikTok, and YouTube. The increased reliance on algorithms raises questions about fairness and transparency in the social media landscape. Pasipamire and Muroyiwa (2024) state that the use of algorithms is a threat to information fairness and trust in AI applications. This reliance also leads to biases, compromises the integrity of information dissemination and undermines public confidence. They argue that inclusive design and community engagement are essential to mitigate bias and foster equitable technology.

In addition, the use of technology such as recommender systems tend to reinforce users' beliefs and preferences thereby creating what are known as filter bubbles and echo chambers. A systematic review by Qazi et al. (2023) describes this as internet customisation that tends to isolate individuals from other opinions and show them selected content. Biases in recommender systems can come from popularity, position, and demographic factors. Moreover, because of such biases, the content users may systematically exclude marginalised voices or amplify stereotypes. Ndaka and colleagues (2025) expose how social media platforms embed gender stereotypes and how AI recommendation algorithms amplify negative gender norms that disproportionately impact women in the African context..

Birhane (2020) examined the Facebook project to create high-resolution population density maps of Africa using computer vision. The company presented the initiative as a project to help governments and NGOs to plan services for their populations but Birhane argues that it amounted to algorithmic colonialism. By assuming authority to map and represent African territories, Facebook exercised power reminiscent of colonial cartography and extracted data for commercial ends. The project was a mirror of earlier colonial rhetoric of "liberating the bottom billion" by connecting the unconnected, while ignoring local epistemologies and reinforcing Western control over digital infrastructure. Such cases show that AI-driven mapping projects on social media platforms can consolidate corporate knowledge over African landscapes without meaningful participation or benefit for local communities.

During Kenya's 2017 elections, the government hired the British data-analytics company, Cambridge Analytica, to support the ruling party's campaign. A study by KICTANet and CIPESA on disinformation in Kenya's political sphere documents how the firm conducted research and developed targeted messaging, including attack advertisements and sponsored posts on Facebook and WhatsApp. Ekdale and Tully (2019). Cambridge Analytica reportedly harvested data from millions of Facebook profiles without consent and used psychographic microtargeting to influence voter perceptions. This incident shows how the company exploited social media algorithms to manipulate electoral outcomes. The report states that what happened created global awareness about how domestic and foreign actors can use social media as a weapon and user data to influence elections. It also shows how vulnerable African democracies are to algorithmically driven disinformation campaigns.

These cases demonstrate that AI-driven social media systems can reveal colonial power dynamics by controlling access, extracting data, and amplifying disinformation. Regulators and technologists should draw on such empirical studies to raise awareness in communities and support local infrastructure, ensuring that recommendation algorithms respect data sovereignty and cultural diversity.

 Conclusion

In conclusion, we acknowledge that Africa is experiencing growth in a thriving ecosystem of Artificial Intelligence (AI) technologies and systems, developed and promoted by both local and global technology players. However, this AI potential is critical to achieving Africa's sustainable development agenda; it also exposes the continent to profound risks of digital and algorithmic colonialism. In particular, the global landscape of AI and data governance reveals deep inequalities; data and algorithms are not neutral, and this is particularly severe in Africa. Specifically, algorithmic colonialism allows technological systems to reinforce historical power imbalances, extracting value from African societies while leaving little room for local control. Further, dominant AI systems reflect masculine world views and Western culture, causing algorithmic bias and discrimination, since AI systems increasingly shape values, identities, and power relations. Again, African countries lack a strong, localized system for data governance and sovereignty; hence, they urgently need localized frameworks to ensure African data is not only protected but also used in ways that benefit local communities.
Thus, Governments across Africa need to craft AI strategies and policy frameworks that move beyond Western-centric ethical frameworks by developing contextually grounded approaches that reflect African epistemologies, cultures, and priorities. This chapter emphasizes the significance of encouraging responsible, inclusive, and pluralistic models of AI development by critically analyzing the power dynamics ingrained in digital platforms and AI infrastructures. Africa can only guarantee that AI becomes a tool for equity and empowerment rather than a means of resurgent marginalization and exploitation by engaging in such reflective, locally oriented interactions. Implementing business models that account for the availability of alternative socio-material worlds to AI and that embrace response-ability is recommended.